\documentclass[12pt]{iopart}

\usepackage{iopams}

\usepackage{graphicx}
\usepackage{txfonts,color}
\usepackage{ulem}
\usepackage{booktabs}
\usepackage{dsfont}

\makeatother

\usepackage{graphicx}
\usepackage{ulem}
\usepackage{txfonts,color}
\usepackage{dsfont}
\usepackage{booktabs} 
\makeatother

\usepackage{iopams}

\begin{document}

\title[Neural-network-based parameter estimation for quantum detection]{Neural-network-based parameter estimation for quantum detection}

\author{Yue Ban$^{1, 2, *}$, Javier Echanobe$^{3}$, Yongcheng Ding$^{4}$, 
Ricardo Puebla$^{5}$ and Jorge Casanova$^{1,6,*}$ }
\address{$^1$ Department of Physical Chemistry, University of the Basque Country UPV/EHU, Apartado 644, 48080 Bilbao, Spain}
\address{$^2$ School of Materials Science and Engineering, Shanghai University, 200444 Shanghai, People's Republic of China}
\address{$^3$ Department of Electricity and Electronics, University of the Basque Country UPV/EHU, Vizcaya, 48940 Spain}
\address{$^4$ International Center of Quantum Artificial Intelligence for Science and Technology (QuArtist) and Department of Physics, Shanghai University, 200444 Shanghai, China}
\address{$^5$ Centre for Theoretical Atomic, Molecular and Optical Physics, Queen's University Belfast, Belfast BT7 1NN, United Kingdom}
\address{$^6$ IKERBASQUE,  Basque  Foundation  for  Science, Plaza Euskadi 5, 48009 Bilbao,  Spain}
\address{$^*$ Corresponding author; Email: ybanxc@gmail.com and jcasanovamar@gmail.com}

\begin{abstract}
Artificial neural networks bridge input data into output results by approximately encoding the function that relates them. This is achieved after training the network with a collection of known inputs and results  leading to an adjustment of the neuron connections and biases. In the context of quantum detection schemes, neural networks find a natural playground. In particular, in the presence of a target (e.g. an electromagnetic field), a quantum sensor delivers a response, i.e., the input data, which can be subsequently processed by a neural network that outputs the target features. In this work we demonstrate that adequately trained neural networks enable to characterize a target with i) Minimal knowledge of the underlying physical model ii) In regimes where the quantum sensor presents complex responses and iii) Under a significant shot noise due to a reduced number of measurements. We exemplify the method with a development for $^{171}$Yb$^{+}$ atomic sensors. However, our protocol is general, thus applicable to arbitrary quantum detection scenarios.
\end{abstract}
	
\maketitle

\section{Introduction}
Machine learning (ML) techniques are nowadays routinely employed in many areas of scientific research, as well as for industrial applications~\cite{Bishop}. The success of ML tools lies in their versatility to tackle multiple and complex problems and in their universal behavior, i.e. their capability to efficiently reproduce any functional dependence between inputs and outputs~\cite{Bishop}. ML techniques have found a fertile ground in physical sciences~\cite{Carleo:19}.  In these disciplines,  ML techniques are opening new avenues to address complex problems, either for classification tasks, parameter regression, or optimisation purposes, as well as to retrieve fundamental physical concepts~\cite{Iten:20,Kim:20,Seif:20}. Quantum physical problems are also included in the range of action of ML, as exemplified by the fast-growing number of successful applications of ML techniques in this realm~\cite{ML-quantum-physics,DasSarma:19}. Among them, we can list the ability of ML to learn quantum measurements~\cite{ML-quantum-measurement1,ML-quantum-measurement2,ML-tomography}, as well as to design quantum gates, unitary transformations and protocols for quantum state preparation~\cite{ML-quantum-states1,ML-quantum-states2,ML-transformation,Ostaszewski:19}. Moreover, ML can be also used for state and entanglement classification~\cite{Giordani:20,Harney:20,Gray:18}, to generate quantum communication protocols~\cite{Wallnofer:20}, identification of Hamiltonians~\cite{ML-Hamiltonian} and to retrieve and model open quantum system dynamics~\cite{Nagy:19,Hartmann:19,Vicentini:19,Yoshioka:19}. See~\cite{ML-quantum-physics} and references therein for a comprehensive review on the distinct applications of ML to quantum physics.

Among the different sub-fields in quantum technologies, quantum sensing~\cite{quantum-sensing} and quantum metrology~\cite{quantum-metrology} have experienced a significant progress, placing them at the forefront of this new generation of technologies harnessing quantum effects. This is primarily a consequence of the progress made during the last decades that has enabled the isolation, control, manipulation, and readout of individual quantum registers~\cite{Dowling:03}. In quantum sensing protocols, the information is encoded in the states (phase and/or populations) of a quantum register which offers unprecedented spatial resolution as this is, typically, of atomic size. As examples of the latter, on the one hand, it is worth mentioning nitrogen-vacancy centers in diamond~\cite{Doherty:13,Dobrovitski:13} that possess exceptional coherence properties even at room temperature, and thus they are well suited to quantum detection of biological samples~\cite{Schirhagl:14,Wu:16, Wang16}. On the other hand, ion-trap technology has been proved also useful in the realm of quantum sensing and quantum metrology~\cite{Warring13,Ca,Yb-PRL2016}. In addition, different kinds of dynamical decoupling pulse sequences \cite{Lidar-concatenated-DD,Floquet-spectroscopy-DD} have led to extended coherence times in quantum systems to enhance the precision of determining unknown parameters with the limit provided by the Quantum Fisher Information \cite{Fisher1,Fisher2}.

So far, different protocols for quantum sensing have been demonstrated in scenarios where the basic coupling mechanism of the sensor and the target can be clearly described whilst, in addition, key target parameters are easily encoded in the harmonic  response of the quantum register~\cite{Maze08, Taylor08}. Nevertheless, quantum sensors significantly deviate from their ideal response as soon as they fail to meet the demanding approximations required by their {\it working regime} (in Sec.~\ref{Ybmodel} we exemplify the latter for the specific scenario of an $^{171}$Yb$^+$ quantum sensor~\cite{Yb-PRL2016, Yb-Nature, Yb-PRL2017,Yb-hyperfine-qubit,Yb-Bayesian}) or due to uncontrollable interactions with environmental agents. Under these general conditions, the quantum sensor leads to a complex response that challenges the identification of target parameters. In this regard, ML techniques may offer an unrivalled tool to extend the performance of quantum sensors to these complex scenarios, while requiring  minimal knowledge of their microscopic description. Therefore, the application of ML to quantum sensing and quantum metrology is  receiving an increasing attention focusing, e.g., on the optimization of adaptive estimation protocols~\cite{ML-adaptive1,ML-adaptive2,ML-adaptive3,ML-adaptive4,ML-adaptive5,ML-adaptive6,ML-adaptive7}, and on the calibration of quantum sensors~\cite{NN-calibration}. 

In this article, we propose a scheme for parameter estimations using a neural network (NN) that takes as input the quantum measurement data obtained by interrogating a quantum sensor.  This scheme allows for an efficient and accurate parameter estimation, only requiring minimal knowledge of the underlying physical model. We illustrate our NN-based parameter estimation strategy for quantum magnetometry employing an atomic-size ${}^{171}$Yb$^{+}$ system~\cite{Yb-PRL2016,Yb-Nature, Yb-PRL2017,Yb-hyperfine-qubit,Yb-Bayesian} aiming to detect both frequency and amplitude of an incident electromagnetic field, in a parameter regime where the quantum register presents a complex response, i.e., not harmonic. Note this represents a departure to the regime in which current experiments with $^{171}$Yb$^+$ are posed~\cite{Yb-PRL2016,Yb-Nature,Yb-PRL2017}. Thus, with the assistance of our method, the scope of quantum detection experiments gets significantly enhanced.  In addition, since our NN is only exposed to the training/validation/test dataset obtained from experiments, the establishment of the NN and its ability of parameter estimation do not depend on the physical model of the system.  Here, we numerically simulate an experimental data acquisition in realistic physical conditions, including shot noise. The average NN prediction accuracy of the relevant parameters is above $97\%$ when inputting data outside the training set.

The article is organized as follows: In Sec.~\ref{NN} we introduce the basic tools for ML and NN that will be employed for parameter estimation. In Sec.~\ref{Ybmodel} we briefly introduce the main ingredients for magnetometry using an atomic-size quantum sensor device consisting of a ${}^{171}{\rm Yb}^+$ ion~\cite{Yb-Nature, Yb-PRL2016, Yb-PRL2017,Yb-hyperfine-qubit}, while the results for NN parameter estimation are presented in Sec.~\ref{results}. In Sec.~\ref{precision}, we show the prediction precision limit given by Quantum Fisher Information and compare the estimation from the Bayesian estimator and our NN. Finally, Sec.~\ref{conclusions} summarizes the main conclusions of the article.

\begin{figure}[]
	\begin{center}
		\scalebox{0.9}[0.9]{\includegraphics{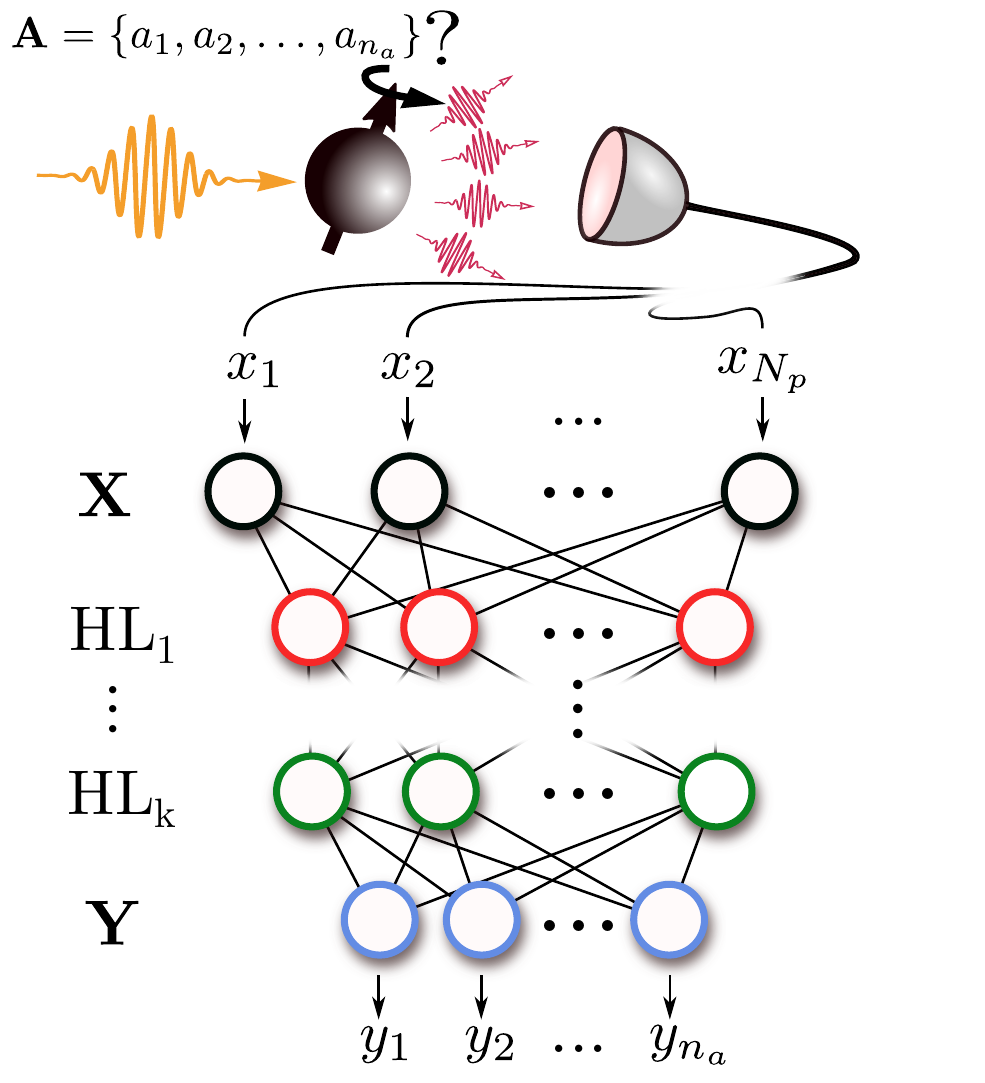}}
		\caption{\label{schematic} Schematic configuration of the quantum parameter estimation employing a NN with $k$ hidden layers, labelled as $\rm{HL}_1$, ..., $\rm{HL}_k$. The quantum sensor interacts with the target signal that triggers a response which is measured. Such quantum sensor response encodes the relevant parameters about the target signal we aim to infer, ${\textbf A}$. The measured data from the quantum register ${\textbf X}$ at different time instances is processed by a number of hidden layers so that the NN finally outputs ${\textbf Y}$, whose  dimension is equal to the number of parameters to be estimated $n_a$. By adequately training the NN with known relations between ${\textbf X}$ and the target parameters ${\textbf A}$, the NN is able to learn the functional dependence $F({\textbf X})={\textbf Y}\approx {\textbf A}$. See main text for details.} 
		\end{center}
\end{figure}

\section{Quantum parameter estimation employing a neural network}\label{NN}
A feed-forward NN can be used to approximate the function that maps inputs to outputs based on the datasets of input-output examples~\cite{Bishop}. During the training process, the parameters of the feed-forward network, i.e. the weights of the connections within layers and the biases of the neurons are optimized iteratively so that the outputs of the network approach the targets within some acceptable errors (cf. Fig.~\ref{schematic} for a schematic representation of the NN). Once the network is properly trained, it enables to estimate the outputs corresponding to the inputs that do not belong to the training set. It is important to note that the NN allows for the training based solely on experimental results, thus only  minimal knowledge of the underlying physical model is required.

We construct a NN such that it takes as inputs the data acquired by measuring the quantum register at  a number $N_p$ of time instants, i.e. the measured data $\textbf{X}=\{x_1,x_2,\ldots,x_{N_p}\}$, and as outputs we demand a number $n_a$ of target parameters to be estimated, denoted by $\textbf{Y}=\{y_1,y_2,\ldots,y_{n_a}\}$. In the training stage, the data $\textbf{X}$ is fed into the NN with its corresponding known output parameters $\textbf{A}=\{a_1,a_2,\ldots,a_{n_a} \}$, so that the weights and neural connections within the hidden layers are optimized to achieve $F({\textbf X})=\textbf{Y}\approx \textbf{A}$ where $F({\textbf X})$ denotes the action of the NN on the input ${\textbf X}$ (cf. Fig.~\ref{schematic})

We define the mean square error $C$ as our cost function  
\begin{eqnarray}
\label{cost-function}
C = \sum_{j=1}^{N} \sum_{i =1}^{n_a} \frac{1}{n_a N} (y_{i}^j - a^{j}_{i})^2,
\end{eqnarray}
for the training  set, where the superscript $j$ accounts for corresponding number among the $N$ examples. The value of $C$ for the validation/test set can be obtained similarly by using the outputs $y_i^j$ from the corresponding  set (i.e. validation and test).
We use the gradient descent algorithm to train our NN, where Levenberg-Marquardt backpropagation is applied as it is usually the fastest one. Since the weights and biases are randomly initialised, we train the network several times to obtain statistical-significance values. By applying backward propagation steps and finding partial derivatives of the weights ${\textbf{w}}$ and the bias ${\textbf{b}}$, i.e., $\partial C/\partial \textbf{w}$ and ${\partial C}/{\partial \textbf{b}}$, these trainable parameters of the NN are adjusted to minimize $C$, expressed as 
\begin{eqnarray}
\label{wb}
\textbf{w} = \textbf{w} - \eta \frac{\partial C}{\partial \textbf{w}}, \quad \textbf{b} = \textbf{b} - \eta \frac{\partial C}{\partial \textbf{b}},
\end{eqnarray}
where $\eta$ stands for the learning rate. 

\begin{figure}[t]
	\begin{center}
		\scalebox{0.3}[0.3]{\includegraphics{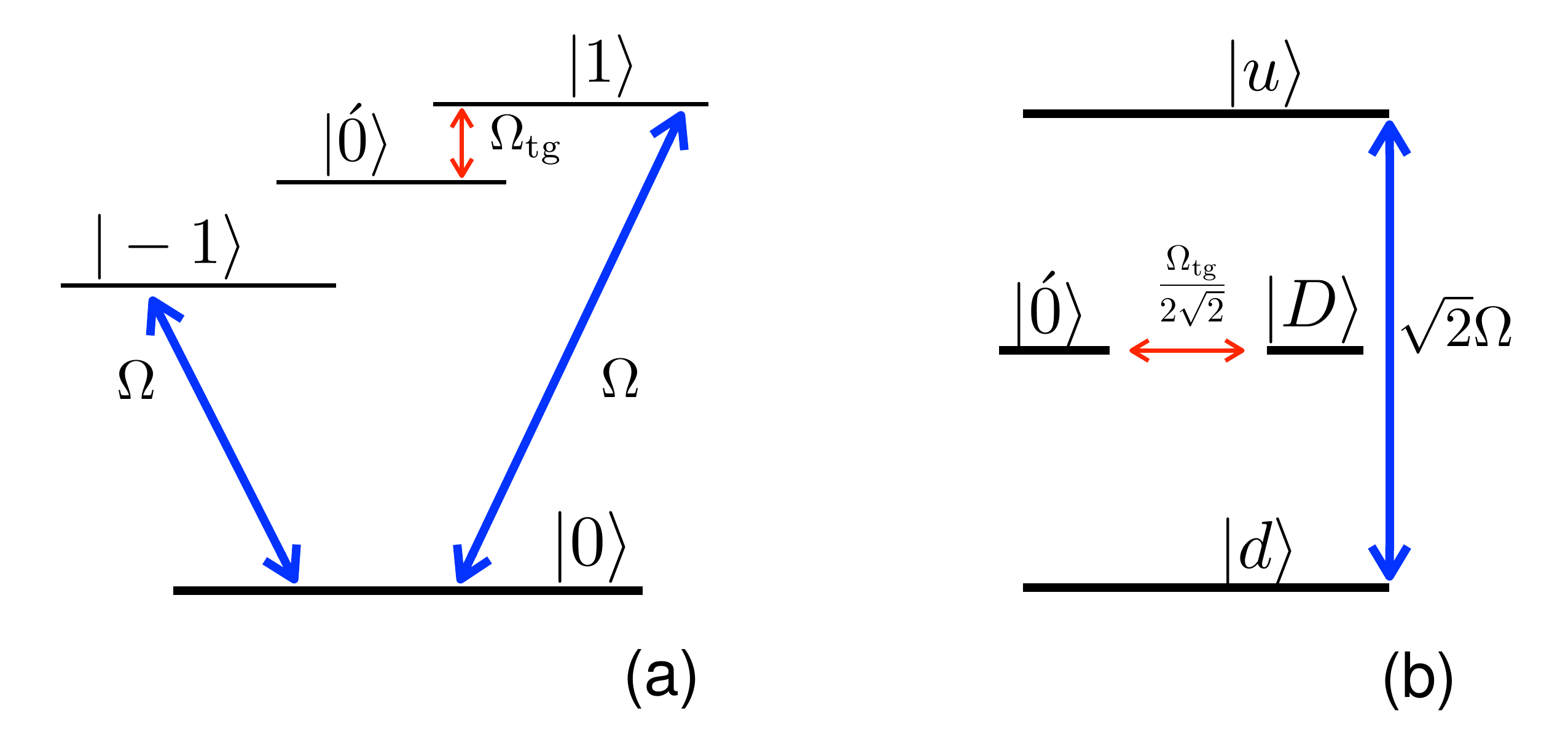}}
		\caption{\label{Ybscheme} (a) Scheme of the relevant energy levels of the $^{171}$Yb$^+$ ion in the presence of a magnetic field where two external MW fields with Rabi frequency $\Omega$ resonantly drive the $|0\rangle \rightarrow |1\rangle$ and $|0\rangle \rightarrow |-1\rangle$ transitions. A target electromagnetic field with amplitude $\Omega_{\rm{tg}}$  can be detected by using the $|\acute{0}\rangle \rightarrow |1\rangle$ state transition. (b) The dressed states of the $^{171}$Yb$^+$ ion, namely, $|u\rangle$, $|d\rangle$, $|\acute{0}\rangle$ and $|D\rangle$, where Rabi oscillation between the states $|D\rangle$ and $|0\rangle$ occurs at a rate $\Omega_{\rm{tg}} / (2\sqrt{2})$.}
	\end{center}
\end{figure}

In this manner, we achieve the parameter estimation scheme.  That is, by introducing the measured data from the quantum register, ${\textbf X}$, to the NN we approximately obtain the unknown target features, i.e. $F({\textbf X})={\textbf Y}\approx {\textbf A}$, in situations where the sensor presents a non-harmonic response. In the following we illustrate the good performance of our strategy using a $^{171}$Yb$^+$ ion as a magnetometer~\cite{Yb-Nature, Yb-PRL2016, Yb-PRL2017,Yb-hyperfine-qubit}.

\section{Case study: Magnetometry with an atomic-size sensor}\label{Ybmodel}

\begin{figure}[b]
	\begin{center}
		\scalebox{0.6}[0.6]{\includegraphics{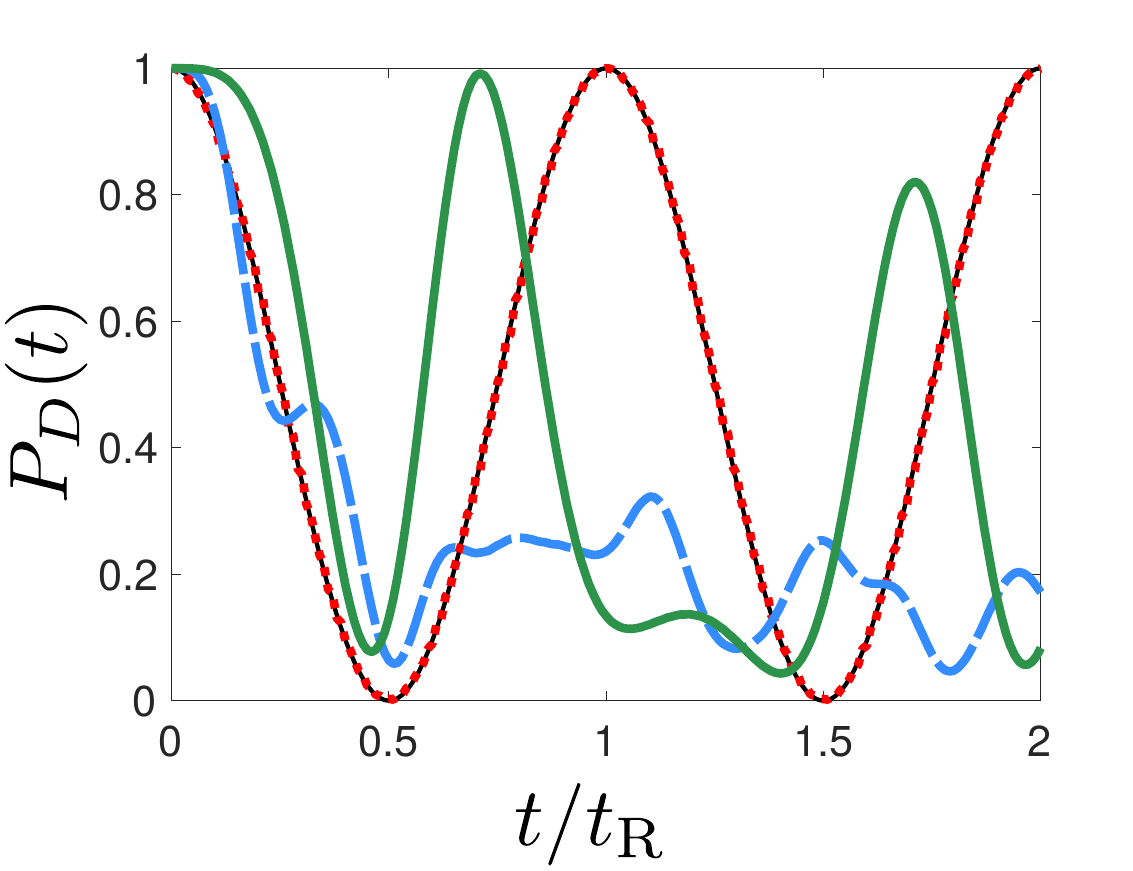}}
		\caption{\label{signals} The time-dependent sensor response  $P_D(t)$ as a function of the evolution time $t/t_{\rm R}$ with $t_{\rm{R}} = 2\pi \sqrt{2}/\Omega_{\rm{tg}}$. The response $P_D(t)$ is obtained by solving the Schr\"{o}dinger equation under $H(t)$ (cf. \ref{app:a}) for three different parameter sets. These are $\Omega_{\rm{tg}}=2\pi\times 1$ kHz, $\xi = 0$, $B_z=1$ mT (dotted red), $\Omega_{\rm{tg}} = 2\pi\times 14$ kHz, $\xi = 2\pi\times 0.3$ kHz, $B_z=1$ mT (dashed blue) and $\Omega_{\rm{tg}} = 2\pi \times 9$ kHz, $\xi = -2\pi \times 0.2$ kHz, $B_z=0.4$ mT (solid green). For zero detuning and  low amplitude, $\Omega_{\rm{tg}}=2\pi\times 1$ kHz, $\xi = 0$, the response (dotted red) overlaps with the ideal harmonic response $P_D(t)=\cos^2(\pi t/t_{\rm{R}})$ (solid black). Other parameters of the system are $P_D(0) = 1$, $\Omega = 2 \pi \times 37.27$ kHz, and $B_z=1$ mT ($B_z=0.4$ mT) corresponding to $\omega_{1}-\omega_{\acute{0}}\approx 2\pi\times 14$ MHz  ($\omega_{1}-\omega_{\acute{0}}\approx 2\pi\times 5.6$ MHz).}
	\end{center}
\end{figure}

Let us consider an $^{171}$Yb$^+$ quantum sensor device aimed to detect electromagnetic fields, according to the scheme put forward in~\cite{Yb-Nature, Yb-PRL2016, Yb-PRL2017}. The sensor is encoded in the $^2S_{\frac{1}{2}}$ manifold of the $^{171}$Yb$^+$ consisting of four hyperfine levels $|0\rangle$, $|\acute{0}\rangle$, $|1\rangle$ and $|-1\rangle$, where the degeneracy of the states $|\acute{0}\rangle$, $|1\rangle$ and $|-1\rangle$ is lifted by applying a static magnetic field $B_z$. In order to cancel the magnetic field fluctuations and achieve long coherence times, we apply two microwave drivings  resonant with the $|0\rangle \leftrightarrow |1\rangle$ and $|0\rangle \leftrightarrow |-1\rangle$ hyperfine transitions with amplitudes $\Omega_{1,2}$ respectively, cf. Fig.~\ref{Ybscheme}. 

A target electromagnetic field of frequency $\omega_{\rm{tg}}$ and amplitude $\Omega_{\rm tg}$, i.e., of the form  $\Omega_{\rm{tg}} \cos(\omega_{\rm{tg}}t)$, can be detected by using the $|\acute{0}\rangle \leftrightarrow |1\rangle$ or $|\acute{0}\rangle \leftrightarrow |-1\rangle$ transition.  Following the scheme proposed in~\cite{Yb-PRL2016}, the target field can be probed assuming that it drives one of the above two transitions, e.g., the $|\acute{0}\rangle \leftrightarrow |1\rangle$ transition with $ \omega_{\rm{tg}} = \omega_1 -\omega_{\acute{0}}+\xi $, where $\xi$ is a small detuning with respect to the resonant condition. In the rotating frame of $\Omega/\sqrt{2} (|u\rangle\langle u| -|d\rangle \langle d|)$ (assuming $\Omega\equiv \Omega_{1,2}$) one finally finds the relevant contributions for the quantum sensor dynamics which is in turn protected against magnetic field fluctuations. 
More specifically, in the dressed state basis $\left\{|u\rangle, |d\rangle, |D\rangle, |\acute{0}\rangle\right\}$  (cf. Fig.~\ref{Ybscheme}) where $|u\rangle = (|B\rangle + |0\rangle) / \sqrt{2}$, $|d\rangle =  (|B\rangle-|0\rangle)/ \sqrt{2}$, $|D\rangle =(|-1 \rangle -|1\rangle)/\sqrt{2}$, $|\acute{0}\rangle = |\acute{0}\rangle$, with $|B\rangle =  (|-1\rangle+|1\rangle)/\sqrt{2}$, we measure the response of the sensor $P_D(t)$. This is, the survival probability of the state $|D\rangle$ when the target electromagnetic field is acting on the sensor.

As we explain in \ref{app:a}, by further assuming that $\xi=0$ and upon a rotating wave approximation ($\Omega_{\rm tg}\ll \Omega\ll \omega_{\rm tg}$) one can find a harmonic/ideal response $P_D(t)= \cos^2(\pi t/ t_{\rm{R}})$, with $t_{\rm{R}} = 2 \pi\sqrt{2}/\Omega_{\rm{tg}}$. At this point, it is worth  mentioning that the conditions  $\xi=0$ and $\Omega_{\rm tg}\ll \Omega\ll \omega_{\rm tg}$ determine the {\it working regime} of the $^{171}$Yb$^+$ quantum sensor leading to its ideal harmonic response. In this situation, the populations of the states $|D\rangle$ and $|\acute{0}\rangle$ oscillate at a rate given by $\Omega_{\rm{tg}} / (2\sqrt{2})$, see \ref{app:a}. 
Remarkably, for a possibly detuned electromagnetic field, $\xi\neq 0$, with a large amplitude $\Omega_{\rm tg}$  and/or small frequency $\omega_{\rm tg}$, the sensor abandons its working regime. 
%
As a consequence, the sensor response $P_D(t)$ no longer follows the ideal harmonic behavior, which challenges the identification of the relevant parameters $\Omega_{\rm tg}$ and $\xi$.

Indeed, a detailed description of the interaction between the internal levels of $^{171}$Yb$^+$ and the target electromagnetic signal (see ~\ref{app:a}) reveals a complex sensor response $P_D(t)$, that needs to be computed by numerically solving the time-dependent Schr{\"o}dinger equation. In Fig.~\ref{signals} we provide three different and illustrative cases for $P_D(t)$. As it can be seen in Fig.~\ref{signals}, as soon as $\Omega_{\rm tg}$ becomes comparable to $\Omega$ and/or $\xi\neq 0$, the response $P_D(t)$ looses its harmonic behaviour. In the following, we show how an appropriately trained NN is able to extract $\Omega_{\rm tg}$ and $\xi$ even in these challenging situations.

\section{Results}\label{results}
\subsection{Setup of the neural network}
We build a NN to estimate a target rf-field characterized by a Rabi frequency $\Omega_{\rm{tg}}$ and its frequency detuning $\xi$ employing a $^{171}$Yb$^+$ ion system. For that, the input layer consisting on $N_p$ neurons takes the array of measured data ${\textbf X}$ (sensor response) where $x_i$ corresponds to the population $P_D(t_i)$ measured after an evolution time $t_i$, while the outputs $\textbf{Y} = \left\{y_1, y_2\right\}$ are expected to approach the targets $\textbf{A} = \left\{a_1, a_2\right\} = \{\Omega_{\rm{tg}},\xi\}$ ($n_a=2$) within a certain error tolerance. We set $N_p=101$ and consider five hidden layers with $40, 20, 12, 6, 3$ neurons, respectively, which are  sufficient to deal with our problem. The activation functions for the hidden layers are of hyperbolic-tangent fashion, while for the output layer a linear behaviour is chosen. These standard activation functions guarantee the good performance of our NN. In the training stage, we minimize the cost function $C$ (cf. Eq.~(\ref{cost-function})) and find that best prediction results are obtained using a learning rate $\eta = 5\cdot 10^{-3}$ in Eq.~(\ref{wb}).

\begin{figure}[]
	\begin{center}
		\scalebox{0.22}[0.22]{\includegraphics{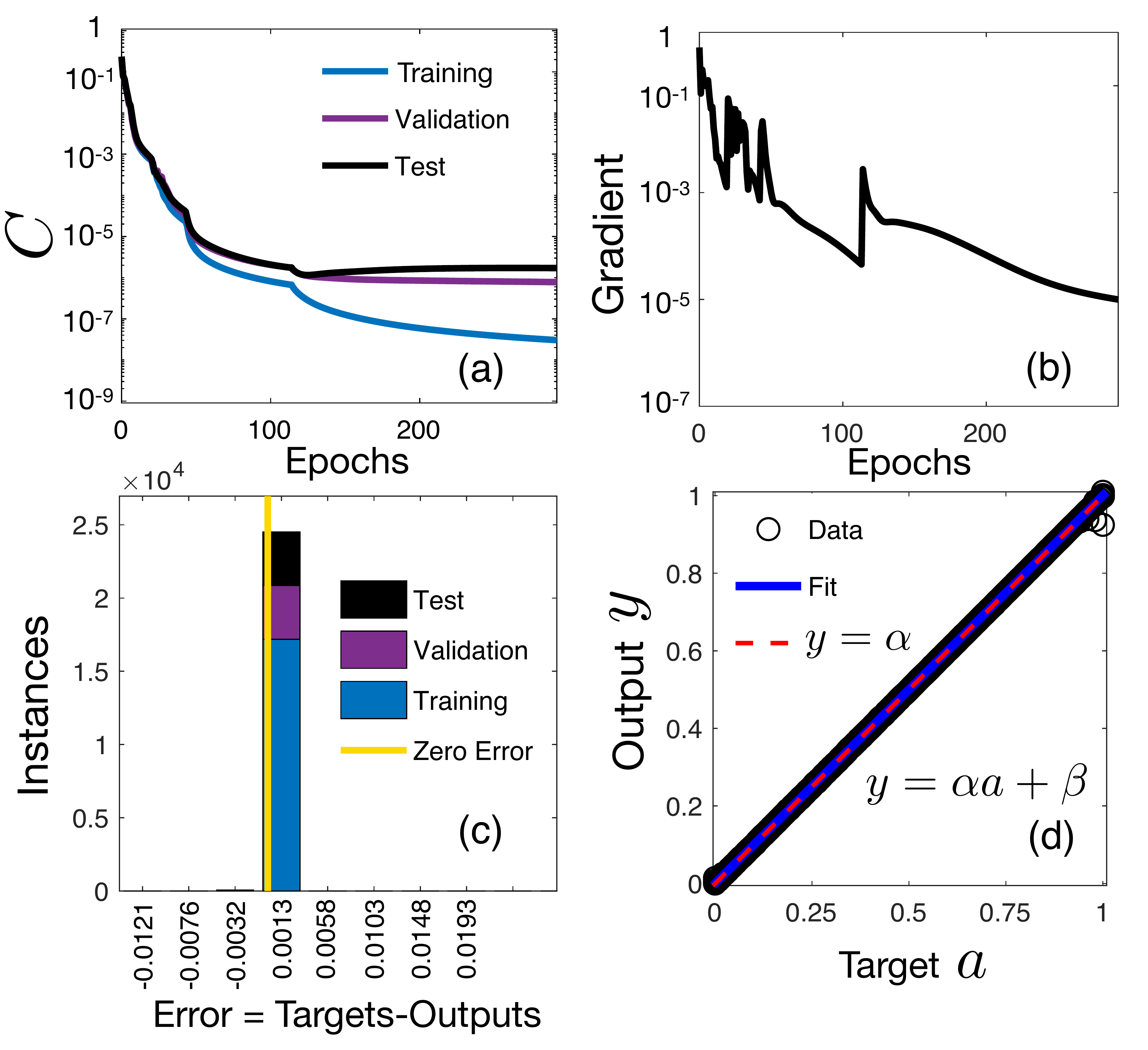}}
		\caption{\label{ideal}  The training results of the NN: (a) Cost function values for training (blue) / validation (purple) / test (black) datasets and (b) gradient of the training set at each epoch. The training stops when the gradient becomes smaller than $10^{-5}$, which is achieved at the $292$th epoch. In panel (c) we show the error histogram at this last epoch, and (d) the comparison between the NN outputs $y=(y_1^r, y_2^r)$ and the targets $a = (\Omega_{\rm{tg}}^r, \xi^r)$. The  input data $\textbf{X}$ is derived from $241$ and $51$ different values for $\Omega_{\rm{tg}}/(2\pi)\in [1,25]$ kHz and $\xi/(2\pi)\in [-0.3,0.3]$ kHz, respectively. The fit in (d) shows the linear relation between two outputs $y=\{y_1^r,y_2^r\}$ and two targets $a=\{\Omega_{\rm{tg}}^r, \xi^r \}$, i.e. $y_1^r=\alpha \Omega_{\rm{tg}}^r  +\beta$ and $y_2^r=\alpha \xi^r  +\beta$ (solid blue), where $1-\alpha \approx 10^{-5} $ while the offset is found to be $\beta \approx 2\cdot 10^{-5}$. As it can be seen, the fit, the linear (ideal) case $y_1^r = \Omega_{\rm{tg}}^r$ and $y_2^r = \xi^r$ (dashed red), and the data almost overlap.} 
		\end{center}
\end{figure}

\subsection{Neural network Magnetometry}

To use the supervised learning strategy, we need to build a training/validation/test set for the NN. Provided by specific sets of $\Omega_{\rm{tg}}$ and $\xi$, the input data $\textbf{X}  = \left\{P_1, P_2, ..., P_{N_p}\right\}$ (with $P_i$ the probability of finding the state $|D\rangle$) is collected at every time instant $t=t_i$ in the time interval $[0.5t_0, t_0]$ where $N_p=101$ and we arbitrarily choose $t_0 = 1.41$ ms. Note that the latter corresponds to one period of the sensor response for the ideal harmonic case when $\Omega_{\rm{tg}} = 2\pi\times 1$ kHz and $\xi=0$.

Before considering a situation with a reduced number of measurements, it is worth highlighting the high accuracy of the estimations using a NN when no shot noise is included. This scenario  can be deemed as the limiting case when a large number of measurements are performed. In order to explore the performance of our NN beyond this ideal scenario, we inspect the range of the parameters $\Omega_{\rm{tg}}/(2\pi) \in  [1, 25]$ kHz and $\xi/(2\pi) \in  [-0.3, 0.3]$ kHz, where the sensor responses clearly deviate from the harmonic (ideal) behavior, as it is shown in Fig. \ref{ideal}. To train our NN with enough examples (this is a sufficiently large number of input data strings $\textbf{X} = \left\{P_1, P_2, ..., P_{N_p} \right\}$, where each $P_i$ is obtained by numerically evolving the system with the total Hamiltonian $H(t)$ in~\ref{app:a}), we derive the examples by extracting $241$ values for $\Omega_{\rm{tg}}$ with the  interval $0.1$ kHz and $51$ values for $\xi$ separated by $0.012$ kHz.  
Therefore, the dataset contains $241\times 51 = 2651$ examples, from  which $70\%$, $15\%$, $15\%$ form the training, validation and test sets. As rescaling input data is a standardized procedure for data processing in a NN, all the input data $\textbf{X}$ are rescaled into the range $[0,1]$. Correspondingly, the targets $\textbf{A}$ are also rescaled into $a= \{\Omega_{\rm tg}^{r}, \xi^r \}$ in the range $[0, 1]$ before being used in the NN. From the outputs $y=\{y_1^r, y_2^r\}$ (also in the same range) obtained from the NN, we can get the results of the estimation $\textbf{Y}$ with the real units.

Fig.~\ref{ideal} shows the training results for the NN, where the training stops when the gradient drops below  $10^{-5}$, which in this case corresponds to the $292$th epoch. In Fig.~\ref{ideal}~(a) we find the performance, i.e., the value of the cost function at each epoch for the training/validation/test set, in Fig. 4~(b) the gradient of the training set at each epoch, at the $292$th epoch, in Fig. 4~(c) the error histogram, and in Fig. 4~(d) the regression. In Fig. 4~(d), the comparison between the outputs of the NN $y_1^r, y_2^r$ and the targets $\Omega^r, \xi^r$ is illustrated. Each circle in the y/a plane corresponds to (i.e. has as coordinates) $(y_1^r, \Omega^r)$ or $(y_2^r, \xi^r)$. The regression of the outputs on the targets is the fit (solid blue line) $y_1^r = \alpha \Omega_{\rm{tg}}^r +\beta$ and $y_2^r = \alpha \xi^r +\beta$ where we get $\alpha\approx 1$ and $\beta\approx 0$. This is, the fits almost coincide with the ideal linear relation (dashed red line) $y_1^r = \Omega_{\rm{tg}}^r$ and $y_2^r = \xi^r$, which indicates that the NN provides outputs approaching the targets with high accuracy.

\begin{figure*}[t]
	\begin{center}
		\scalebox{0.28}[0.28]{\includegraphics{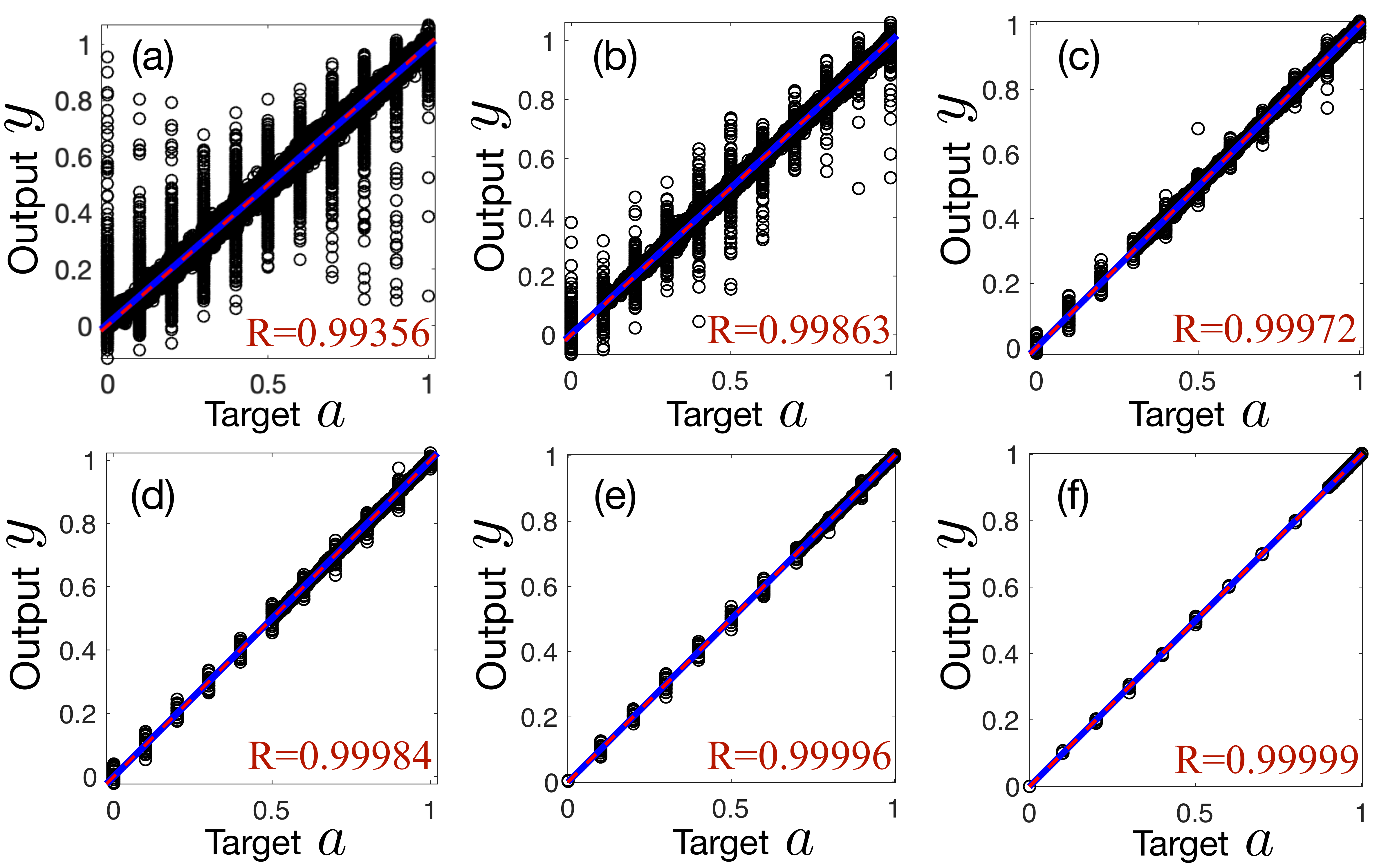}}
		\caption{\label{regression-shotnoise} Comparison between the outputs of the NN $y = (y_1^r, y_2^r)$ and the targets $a= (\Omega^r, \xi^r)$ including shot noise, with different range of $\Omega_{\rm{tg}}^r$, namely: (a) $\Omega_{\rm tg}^r\in[0, 1]$, (b) $[0.1, 1]$, (c) $[0.3, 1]$, (d) $[0.5, 1]$, (e) $[0.7, 1]$ and (f) $[0.9, 1]$ while keeping the same detuning range $\xi^r\in [0,1]$ (i.e., $\xi/(2\pi)\in [-0.3, 0.3]$ kHz). In all plots the fit $y_1^r = \alpha \Omega_{\rm{tg}}^r +\beta$ and $y_2^r = \alpha \xi^r +\beta$ (solid blue) are  almost equivalent to the ideal dependences $y_1^r = \Omega_{\rm{tg}}^r$ and $y_2^r = \xi^r$  (dashed red) for both parameters and (a) $1-\alpha \approx 10^{-2}$, $\beta \approx 6\cdot 10^{-3}$, (b) $1-\alpha \approx 4\cdot 10^{-3}$, $\beta \approx 3\cdot 10^{-3}$, (c) $1-\alpha \approx 7\cdot 10^{-4}$, $\beta \approx 3\cdot 10^{-4}$, (d) $1-\alpha \approx 3\cdot 10^{-4}$, $\beta \approx 2\cdot 10^{-4}$, (e) $1-\alpha \approx 10^{-4}$, $\beta \approx 4\cdot 10^{-5}$, (f) $1-\alpha \approx 7\cdot 10^{-5}$, $\beta \approx 2\cdot 10^{-5}$. See main text for further details. 
		
		} 
	\end{center}
\end{figure*}

\begin{figure}[t]
	\begin{center}
		\scalebox{0.41}[0.41]{\includegraphics{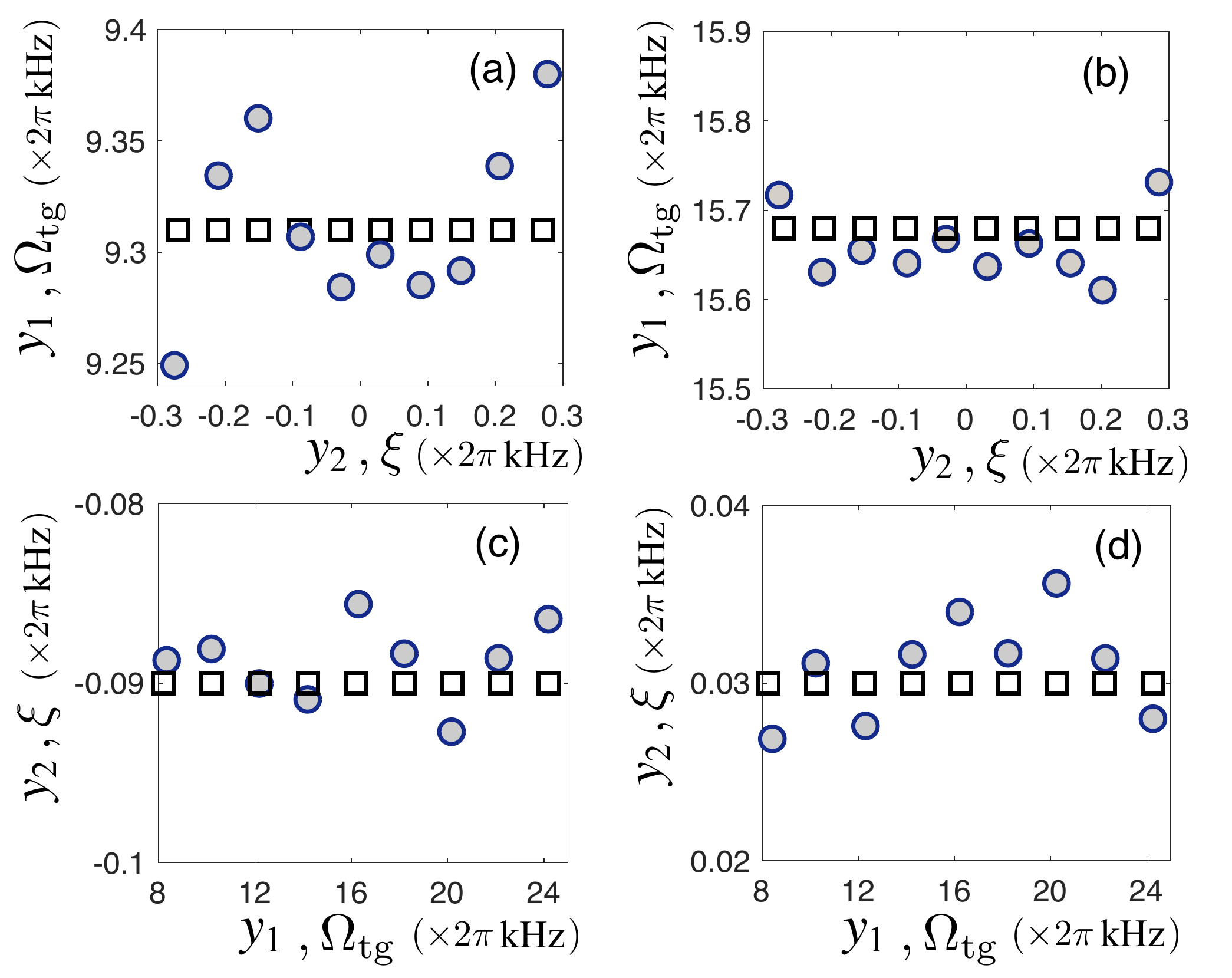}}
		\caption{\label{test-results} Comparison of the estimation results ${\textbf Y}=\{y_1,y_2\}$ (full circles) with respect to their the target values ${\textbf A}=\{a_1,a_2\}=\{\Omega_{\rm tg},\xi\}$ (open squares). The outputs $y_1$ and $y_2$ are obtained from the NN trained with the dataset using $\Omega_{\rm{tg}}/(2\pi) \in [8.2, 25]$ kHz and $\xi/(2\pi) \in  [-0.3, 0.3]$ kHz. The inputs, i.e. the sensor response, are derived from randomly chosen targets $\xi/(2\pi) \in [-0.27, -0.21, -0.15, -0.09, -0.03, 0.03, 0.09, 0.15, 0.21, 0.27]$ kHz when $\Omega_{\rm{tg}} = 2\pi \times 9.31$ kHz in (a), and $\Omega_{\rm{tg}} = 2\pi \times 15.68$ kHz in (b). The NN also works for the sensor response obtained from $\Omega_{\rm{tg}}/(2\pi) \in  [8.25, 10.25, 12.25, 14.25, 16.25, 18.25, 20.25, 22.25, 24.25]$ kHz when $\xi = -2\pi\times0.09$ kHz in (c), and $\xi = 2\pi \times 0.03$ kHz in (d). Every input is simulated as an experimental acquisition with the measurement of $N_m=100$ shots at each time instant $t_i$.}
	\end{center}
\end{figure}

In situations with a significant shot noise we have to feed the NN with data including potential statistical fluctuations. In order to generate the input data with shot noise, we numerically simulate an experimental acquisition to include shot noise as follows. At every time instant $t=t_i$, the  measurement results in a binary outcome $z_{n;i} \in \{0, 1\}$ at each shot. Consequently, the measurement result is $P_i = \sum_{n=1}^{N_m} z_{n;i}/N_m$ where $z_{n;i}$ are drawn from a Bernoulli distribution, this is $z_{n;i} \sim B(1, P_D(t_i))$, and $N_m$ is the number of shots. The statistical fluctuation in $P_i$ due to the finite number of measurements is reduced when the number of measurements $N_m$ increases. 

In the following we consider each shot measurement for obtaining the expectation value $P_i$ with $N_m =100$.  We generate $ 241 \times 11 \times 20$ examples to build a training/validation/test dataset, where $241$ and $11$ refer to the number of distinct values for $\Omega_{\rm{tg}}$ and $\xi$, respectively. Due to the existence of shot noise, the expectation value $P_i$ at every $t_i$ varies for each experimental realization (i.e. for the same $\Omega_{\rm{tg}}$ and $\xi$). To include this statistical fluctuations in the NN, we perform $20$ repetitions for each of the simulated experimental acquisitions.  Thus, we collect more than one example for every available target, which helps the NN to find reliable estimations. We find that the dataset with $20$ different repetitions during the time interval $[0.5 t_0, t_0]$ are enough for our NN to learn the relation between the sensor response $\textbf{X} = \{P_1,\ldots,P_{N_p}\}$ and the targets ${\textbf A}=\{\Omega_{\rm tg},\xi\}$ in the presence of shot noise.

We divide the dataset in $241 \times 11 \times 20$ examples according to the following intervals: $\Omega_{\rm{tg}}/(2\pi) \in  [1, 25]$ kHz, $[3.4, 25]$ kHz, $[8.2, 25]$ kHz, $[13, 25]$ kHz, $[17.8, 25]$ kHz, and $[22.6, 25]$ kHz, which correspond to rescaled values $\Omega_{\rm tg}^r\in[0,1]$, $[0.1, 1]$, $[0.3, 1]$, $[0.5, 1]$, $[0.7, 1]$, $[0.9, 1]$, respectively, while $\xi$ is kept in the same range as before, namely, $\xi/(2\pi)\in [-0.3, 0.3]$ kHz. Using the same cost function as Eq.~(\ref{cost-function}) and extracting $70\%$, $15\%$, $15\%$ of examples in each dataset mentioned above to create the training/validation/test set, respectively, we train each NN based on each dataset. The results are summarized in Fig.~\ref{regression-shotnoise}.  On the one hand, in all subplots where each circle refers to the point $(y_1^r, \Omega^r_{\rm{tg}})$ or $(y_2, \xi^r)$, the linear regression relation $y_1^r = \alpha \Omega_{\rm{tg}}^r +\beta$ and $y_2^r = \alpha \xi^r +\beta$ between the outputs $y=\{y_1^r, y_2^r\}$ and the targets $a=\{\Omega^r_{\rm{tg}}, \xi^r\}$ is compatible with the expected linear relation $y_1^r = \Omega_{\rm{tg}}^r$ and $y_2^r= \xi^r$. In all the cases, the correlation coefficient $\rm{R}$ is higher than $99\%$. Note that $\rm{R}$ measures the linear dependence between the outputs and the targets (the correlation coefficient and the fit line are taken directly from MATLAB). 
Therefore, the NN successfully identifies the target features, namely, $F({\textbf X})={\textbf Y} \approx {\textbf A}$ to a very good approximation. 
On the other hand, the deviation between the two outputs and their corresponding targets in the training/validation/test sets gets smaller for NN trained using large values of $\Omega_{\rm{tg}}$. The fit line $y = \alpha a +\beta$ approaches the expected linear relation $y = a$, when smaller $\Omega_{\rm tg}$ values are excluded, as $\alpha$ and $\beta$ tend to $1$ and $0$ progressively (see Fig. \ref{regression-shotnoise}  from (a) to (f)).  In particular, we find that under our considerations, our NN performs exceptionally good, this is $\rm{R} = 0.99972$,  for the dataset with $\Omega_{\rm{tg}} >2\pi \times 8.2$ kHz, while the estimation accuracy slightly drops ($\rm{R} = 0.99356$) when considering the whole range $\Omega_{\rm tg}/(2\pi)~\in [1,25]$ kHz.  This is a consequence of the estimation of the parameter $\xi$: Owing to the nature of the sensor response, smaller values of $\Omega_{\rm tg}$, more difficult the identification of $\xi$. Yet, as we discuss later, one can still achieve similarly accurate NN-based estimations for these cases by interrogating the quantum sensor after a longer evolution time.

We pick up the NN trained from the dataset with the range $\Omega_{\rm{tg}}/(2\pi) \in [8.2, 25]$ kHz and $\xi/(2\pi) \in  [-0.3, 0.3]$ kHz. To prove the high accuracy of the estimations using the trained NN, we randomly choose target parameters outside the training/validation/test dataset and check the outputs from the NN. The average value of the accuracy $F_1=\frac{1}{N}\sum\nolimits_{j=1}^N |y_1^j-\Omega^j_{\rm{tg}}|/\Omega^j_{\rm{tg}}$ and  $F_2 =\frac{1}{N} \sum\nolimits_{j=1}^N |y_2^j -\xi^j|/\xi^j$ reveal that with randomly chosen $N =38$ examples, the predictions coincide with the targets to a very good approximation ($F_1>99.8\%$, $F_2 > 97\%$), as illustrated in Fig.~\ref{test-results}.

As discussed in \ref{app:b}, we find that shot noise impacts the estimation of the parameters in a different fashion. Indeed, we find that it is  more difficult for the NN to learn $\xi$ values at small $\Omega_{\rm{tg}}$, while the prediction on $\Omega_{\rm{tg}}$ has higher accuracy.  At small $\Omega_{\rm{tg}}$, shot noise hinders the unequivocally identification of the sensor response with their corresponding target values $\Omega_{\rm tg}$ and $\xi$. As an example, for $\Omega_{\rm tg}\approx 2\pi\times 1$ kHz, the NN is unable to distinguish between the different values of $\xi$. In this situation, if one slightly tunes $\xi$, the fluctuation of shot noise exceeds the difference of the ideal sensor response obtained from the theoretical model at short times (see \ref{app:b}). 
Meanwhile, we also find that using $11$ examples of $\xi$ is enough to establish a NN trained from the dataset with the range $\Omega_{\rm{tg}}/(2\pi) \in [8.2, 25]$ kHz and $\xi/(2\pi) \in  [-0.3, 0.3]$ kHz, and it works well to predict parameters.

It is possible to improve the detection scheme by allowing for a longer evolution time at the cost of time and efforts. In this manner, the previously indistinguishable sensor responses can be individually identified even in presence of shot noise fluctuations (cf. \ref{app:b}). Therefore, collecting the training/validation/test dataset at longer time could obviously be an option to obtain a NN which has better abilities to estimate the parameters close to the resonant case, i.e. $\xi\approx 0$. For example, one can build up such a dataset at the interval $[2.5t_0, 3 t_0]$, so that the measured sensor response with $\xi=2\pi\times 0.12$ kHz and $\xi =2\pi\times 0.06$ kHz for $\Omega_{\rm{tg}}=2\pi\times 1$ kHz can be distinguished (cf. \ref{app:b}). Clearly, a higher accuracy for $\xi$ demands a longer evolution time. As an example we take again $\Omega_{\rm tg}=2\pi\times 1$ kHz, but aiming to distinguish between sensor responses with closer values of $\xi$, e.g. $\xi=2\pi\times 0.072$ kHz and $\xi =2\pi\times 0.06$ kHz (cf. Fig.~\ref{signals-shotnoise} in \ref{app:b}). On the one hand, such high accuracy comes with an increased time cost to generate the dataset. On the other hand, larger measurements $N_m$ will reduce the shot noise, thus bringing the measured data closer to the ideal expectation value at the expense of increasing the experimental resources. 

\section{Discussion on the measurement precision}\label{precision}
Quantum fisher Information (QFI), constraining the achievable precision in statistical estimation of the parameter $\theta$, for a pure state $|\psi\rangle$ is defined as  
\begin{eqnarray}
\label{QFI}
I_\theta = 4 \left[ \langle\partial_\theta \psi |\partial_\theta \psi \rangle - | \langle \psi | \partial_\theta \psi \rangle |^2 \right],
\end{eqnarray}
where in our scenario $\theta = \Omega_{\rm{tg}}$ or $\xi$.
Taking the variance for the parameter estimator encoded in a state at the final time instant $t_0$ at which we measure $P_D$, we find that the precision is upper bound, or similarly, the variance is lower bound as
\begin{eqnarray}
\label{bound}
\Delta^2 \theta \ge \Delta^2 \theta(t_0)^{\rm{QFI}} = \frac{1}{N_T I_{\theta}(t_0)}.
\end{eqnarray}
where 
$N_T = N_p\times N_m$ with $N_p=101$ the time points at which the state is interrogated, and $N_m =100$ measurements of shots per time instant. 
For instance, in the case of a harmonic response $\Omega_{\rm{tg}} = 2\pi\times 1$ kHz and $\xi=0$, we can express the state of the system analytically as
\begin{eqnarray}
\label{state}
|\psi(t)\rangle = \cos \left(\frac{\Omega_{\rm{tg}}t}{2\sqrt{2}}\right) |D\rangle + i \sin \left(\frac{\Omega_{\rm{tg}}t}{2\sqrt{2}}\right) |\acute{0}\rangle,
\end{eqnarray} 
From Eq. (\ref{QFI}), we can directly compute $I_{\Omega_{\rm{tg}}} = t^2/2$. Consequently, the variance for the estimator of $\Omega_{\rm{tg}}$ should be equal or larger than $\Delta \Omega_{\rm{tg}}^{\rm{QFI}} \approx 2\pi \times 1.5\cdot 10^{-3}$ kHz.
For the case of non-harmonic responses, we take the instance $\Omega_{\rm{tg}} = 2\pi \times 9.31$ kHz and $\xi=2\pi\times 0.15$ kHz and numerically derive $\Delta \Omega_{\rm{tg}}^{\rm{QFI}} \approx 2\pi \times 3\cdot 10^{-3}$ kHz and $\Delta \xi^{\rm{QFI}} \approx 2\pi \times 7\cdot 10^{-4}$ kHz. 
Now, we generate $100$ examples in the presence of shot noise from the same targets (i.e. $\Omega_{\rm{tg}} = 2\pi \times 9.31$ kHz and $\xi=2\pi \times 0.15$ kHz) and obtain the outputs of the NN  trained from the dataset with the range $\Omega_{\rm{tg}}/(2\pi) \in [8.2, 25]$ kHz and $\xi / (2\pi) \in [-0.3, 0.3]$ kHz. Consequently, we derive the average value $\overline{y}_1 = 2\pi\times 9.3055$ kHz and $\overline{y}_2 = 2\pi \times 0.1523$ kHz with the standard deviation $\sigma_{y_1} = 2\pi\times 0.0339$ kHz and $\sigma_{y_2} = 2\pi \times0.0027$ kHz, which indicates that the precision is below the limit imposed by the QFI. 

It is interesting to note that the obtained precision should always be lower than this upper bound for any kind of estimator.  Here, we compare the prediction of a NN and a Bayesian estimator which is known to be optimal for large datasets. In particular, we compute the posterior distribution following the well-known Bayes theorem, $p(\theta|{\textbf X})\propto p({\textbf X}|\theta)p(\theta)$ where $p(\theta)$, $p({\textbf X}|\theta)$ and $p(\theta|{\textbf X})$ denote the prior, likelihood and posterior distributions. Note that ${\textbf X}$ represents the data obtained by interrogating the quantum sensor at different time instances used to feed the NN. In this case we consider an uninformative prior ($p(\theta)\propto 1$) for the two parameters, while the likelihood is assumed to obey a Gaussian distribution since $N_m\gg 1$, that is, $p({\textbf X}|\theta)=\Pi_{j=1}^{N_p}\frac{1}{\sqrt{2\pi}\sigma_j}e^{-(x_j-\tilde{x}_j)^2/(2\sigma_j^2)}$ where $\tilde{x}_j$ refers to the expectation value of the measured observable, $P_D$ at time $t_j$ under the parameters $\theta$ and computed following the Hamiltonian in Eq.~(\ref{H}) (see~\ref{app:a}). As standard deviation we take $\sigma_j=1/\sqrt{N_m}$. From $p(\theta|{\textbf X})$ one can obtain the estimator for $\theta_j$ and its variance as $\Delta^2\theta_j=\int d\theta (\theta_j-\overline{\theta}_j)^2p(\theta|{\textbf X})$ and $\overline{\theta_j}=\int d\theta \theta_j p(\theta|{\textbf X})$, where integral is performed over the unknown parameters $\theta$. In the example considered above, $\Omega_{\rm tg}=2\pi \times 9.31$ kHz and $\xi=2\pi\times 0.15$ kHz, a Bayesian analysis leads to $\overline{\Omega}_{\rm tg}=2\pi \times 9.31(2)$ kHz and $\overline{\xi}=2\pi \times 0.153(7)$ kHz with $\overline{\Delta \Omega_{\rm tg}}\approx 2\pi \times 2\cdot 10^{-2}$ kHz and $\overline{\Delta \xi}\approx 2\pi \times 7\cdot 10^{-3}$ kHz, which reveals a comparable precision as that obtained with NN. Similar values for the variances are obtained for different datasets and parameters. 
We remark that, a Bayesian analysis and a NN provide similar precision for the estimators. However, a Bayesian estimator requires a precise microscopic model to compute the evolution for each possible combination of the unknown parameters $\theta$. In this regard, a NN needs less knowledge of the system due to the fact that the training/validation/test datasets only originate from the experimental measurement and the input-output relation is \textit{learned}, while at the same time allowing for good estimators. Thus, a well-trained NN can accurately provide estimators for the unknown parameters in a wide range set by the training dataset with a minimal knowledge of the underlying physical model.

\section{Conclusion}\label{conclusions}
We have proposed a scheme for quantum parameter estimation using neural networks (NNs) which can effectively reproduce the functional dependence between the input measured data from the quantum register and the target parameters to be estimated. We have illustrated our scheme for magnetometry using an atomic-size sensor encoded in the internal levels of a $^{171}$Yb$^+$ ion and in a parameter regime where the sensor presents a complex response, far from an ideal harmonic behaviour. The NN allows for an accurate estimation of the amplitude and detuning of the target electromagnetic field for cases  which do not belong to the training dataset. Since ML techniques require  minimal knowledge of the underlying physical model and are able to tackle complex input-output relations, these tools are best placed to enhance the performance of quantum sensors. We expect that our results will motivate further research and applications of NN in quantum sensing and quantum metrology.

\section*{Acknowledgments}
We acknowledge financial support from Spanish Government via PGC2018-095113-B-I00 (MCIU/AEI/FEDER, UE), Basque Government via IT986-16, as well as from QMiCS (820505) and OpenSuperQ (820363) of the EU Flagship on Quantum Technologies, and the EU FET Open Grant Quromorphic (828826). R. P. acknowledges the support by the SFI-DfE Investigator Programme (grant 15/IA/2864). J. C. acknowledges the Ram\'on y Cajal program (RYC2018-025197-I) and the EUR2020-112117 project of the Spanish MICINN, as well as support from the UPV/EHU through the grant EHUrOPE.

\clearpage

\appendix

\section{Details of the atomic-size $^{171}$Yb$^+$ magnetometer}\label{app:a}
The quantum sensor device of the $^{171}$Yb$^+$ ion is encoded in $^2S_{\frac{1}{2}}$ manifold consisting of four hyperfine levels $|0\rangle$, $|\acute{0}\rangle$, $|1\rangle$ and $|-1\rangle$ (See Fig.~\ref{Ybscheme} (a)). As commented in the main text, leading-order magnetic field fluctuations are cancelled by applying two microwave drivings with amplitude $\Omega\equiv \Omega_{1,2}$. These are represented in Fig.~\ref{Ybscheme} (a) with two blue lines. Following the scheme in Ref.~\cite{Yb-PRL2016}, one can find that a target electromagnetic field with amplitude $\Omega_{\rm tg}$ and frequency tuned close to the resonance, $\omega_{\rm tg}=\omega_1-\omega_{\acute{0}}+\xi$ with $\xi$ a small detuning, leads to the harmonic sensor response $P_D(t) = \cos^2(\Omega_{\rm{tg}} t/ 2\sqrt{2})$ by assuming $\Omega_{\rm tg}\ll \Omega\ll \omega_{\rm tg}$, $\xi=0$, and having the initial state prepared in the dressed state $|D\rangle$.  Yet, a more  realistic model without making those approximations leads to a departure from the ideal response. 

In the following we provide a derivation of a Hamiltonian valid for $\xi\neq 0$ and for amplitudes $\Omega_{\rm tg}$ comparable to $\Omega$.
The Hamiltonian of a $^{171}$Yb$^+$ ion in a magnetic field on the $z$ direction, as well as under the effect of a number of MW drivings (labelled with $j$) is~\cite{Yb-hyperfine-qubit}
\begin{equation}
H = A \ {\bf J} \cdot {\bf I}  + \gamma_e B_z J_z - \gamma_n B_z I_z + \sum_j(\gamma_e B_x^j  J_x - \gamma_n B_x^j I_x) \cos{(\omega_j t + \phi_j)},
\end{equation}
where $A \approx (2\pi) \times 12.643$ GHz, $\gamma_e = (2\pi)\times 2.8024$ MHz/G,  $\gamma_n\equiv \gamma_{^{171}\rm Yb^{+}} = (2\pi)\times 4.7248$ kHz/G, while ${\bf J}$ and ${\bf I}$ are spin-1/2 operators in a basis $\{ |1 1\rangle, |1 0\rangle, |0 1\rangle, |0 0\rangle\}$ such that $J_z |0 \ m\rangle  =-\frac{1}{2} |0 \ m\rangle$, $J_z  |1 \ m\rangle =\frac{1}{2} |0 \ m\rangle$ and $I_z |m \ 0\rangle =-\frac{1}{2}|m \ 0\rangle$, $I_z |m \ 1\rangle =\frac{1}{2}|m \ 1\rangle$ for $m=0,1$. 
Now, in a new basis  $\left\{ |1 \rangle, |\acute 0\rangle, |-1\rangle, | 0\rangle\right\}$  that diagonalize $A \ {\bf J} \cdot {\bf I}  + \gamma_e B_z J_z - \gamma_n B_z I_z$ we get
\begin{eqnarray}
H &=& \omega_1 |1\rangle\langle 1| + \omega_{\acute 0} |\acute 0\rangle\langle \acute 0| + \omega_{-1} |-1\rangle\langle -1| + \omega_{0} | 0\rangle\langle 0| \nonumber\\
&+&\sum_j \frac{B_x^j}{2}  \bigg[ c_{1 \acute0} |1\rangle\langle \acute{0}| +  c_{1 0} |1\rangle\langle 0| + c_{\acute 0 -1} |\acute0\rangle\langle -1| + c_{ 0 -1} |0\rangle\langle -1|  + {\rm H.c.} \bigg]   \cos{(\omega_j t +\phi_j)} \nonumber\\
\end{eqnarray}
The first line of the previous equation  leads to the energy scheme in Fig.~\ref{Ybscheme} (a) with $ \omega_{1} = \frac{A}{4} + (\gamma_e-\gamma_n) \frac{B_z}{2}$, $\omega_{-1} = \frac{A}{4} - (\gamma_e-\gamma_n) \frac{B_z}{2}$, $ \omega_{\acute{0}} \approx \frac{A}{4}  + \frac{(\gamma_e+\gamma_n)^2}{4 A} B_z^2$, and $ \omega_{0} \approx -\frac{3A}{4}  - \frac{(\gamma_e+\gamma_n)^2}{4 A} B_z^2$  while each $c_{k,l}$ coefficient can be found by projecting the $J_x$ and $I_x$ operators in the $\left\{ |1 \rangle, |\acute 0\rangle, |-1\rangle, | 0\rangle\right\}$ basis. Note that, in our case $j$ ranges from 0 to 3 as we are including two driving fields (j=1,2) to decouple the sensor from noise, as well as the target field (j=3) that we want to characterize.
Now, in the dressed state basis $\left\{|u\rangle, |d\rangle, |D\rangle, |\acute{0}\rangle\right\}$ (see main text for a specific definition of these states),  and assuming we use the $ | \acute0\rangle \leftrightarrow | 1\rangle$ transition to couple with the target field we get

\begin{eqnarray}
\label{H}
\nonumber
H(t) &=&  \frac{\Omega}{\sqrt{2}} (|u\rangle \langle u| - |d\rangle \langle d|)+\left[\frac{\Omega}{4} (|u\rangle \langle \acute{0}| + |d\rangle \langle \acute{0}| )  - \frac{\Omega_{\rm{tg}}}{2 \sqrt{2}} |D\rangle \langle \acute{0}|\right] e^{- i \xi t} + \rm{H.c.}
 \\
\nonumber
&& - \left[ \frac{\Omega}{2\sqrt{2}} (|u\rangle \langle u| - |d\rangle \langle d|) + \frac{\Omega}{4}(|u\rangle \langle D| +|D\rangle \langle d|) - \frac{\Omega}{4}(|D\rangle \langle u| + |d\rangle \langle D|)
\right] e^{i \gamma_e B_z t} + \rm{H.c.}
\nonumber
\\
&& + \frac{\Omega_{\rm{tg}}}{2} \left(\frac{1}{2} |u\rangle\langle \acute{0}|  + \frac{1}{2} |d\rangle \langle \acute{0}| - \frac{1}{\sqrt{2}} |D\rangle \langle \acute{0}|\right) e^{2i (\frac{\gamma_e B_z}{2}-\frac{\gamma_e^2 }{4A} B^2_z)t} e^{i \xi t} +\rm{H.c.}
\nonumber
\\
&& + \frac{\Omega_{\rm{tg}}}{2} \left(\frac{1}{2} |\acute{0}\rangle \langle u|+ \frac{1}{2} |\acute{0}\rangle \langle d| + \frac{1}{\sqrt{2}}|\acute{0}\rangle \langle D| \right) e^{i\gamma_e B_z t} e^{i \xi t} + \rm{H.c.}
\nonumber
\\
&& + \frac{\Omega_{\rm{tg}}}{2} \left(\frac{1}{2} |\acute{0}\rangle \langle u|+ \frac{1}{2} |\acute{0}\rangle \langle d| + \frac{1}{\sqrt{2}}|\acute{0}\rangle \langle D| \right) e^{i \frac{\gamma^2_e}{2A}B^2_z t} e^{-\xi t} +\rm{H.c.}.
\end{eqnarray}
Here we are considering that the driving fields have $\omega_1 = \omega_1 - \omega_0$, $\omega_2  = \omega_{-1} - \omega_0$, $\phi_{1}=\pi$ and $\phi_2=0$, while the target field departs from the harmonic regime of the sensor with a detuning $\xi \neq 0$ as well as with a Rabi frequency $\Omega_3 \equiv \Omega_{\rm tg}$ such that it can also lead to transitions to the $|-1\rangle$ level, see Fig.~\ref{Ybscheme} (a).

The population $P_D(t)$, i.e. the sensor response, follows from solving the quantum dynamics under the evolution of the previous Hamiltonian, where we have neglected magnetic-field fluctuations as they are not important for the time-scale considered here~\cite{Yb-Bayesian}. The other parameters appearing in the Hamiltonian are the electronic/nuclear gyromagnetic ratio $\gamma_{e/n}$,  two MW drivings with the same amplitude $\Omega = 2 \pi \times 37.27 $ kHz, and the magnetic hyperfine constant $A \approx 2 \pi \times 12.643$ GHz as measured in Ref.~\cite{Yb-hyperfine-qubit}.

\begin{figure*}[t]
	\begin{center}
		\scalebox{0.55}[0.55]{\includegraphics{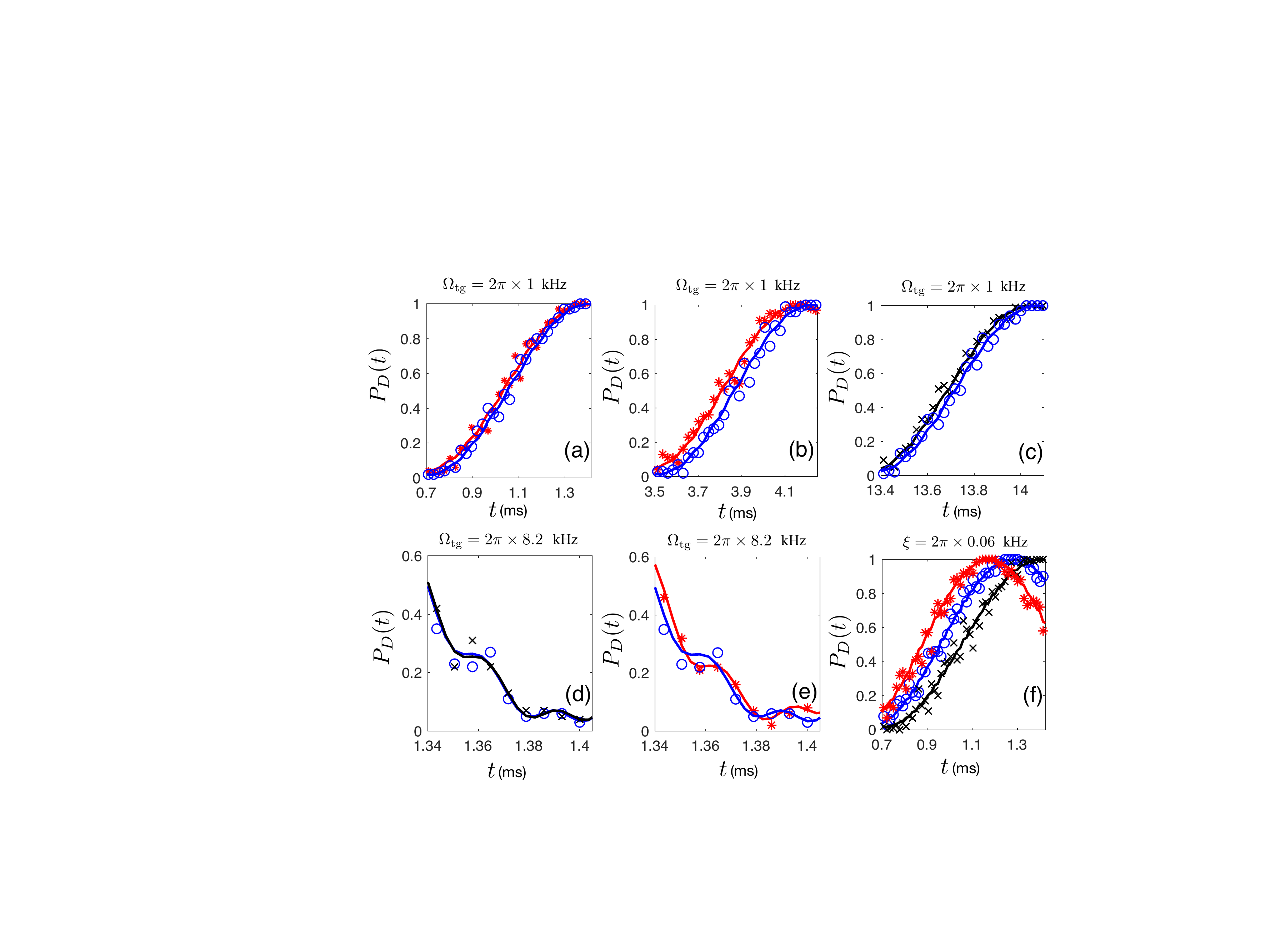}}
		\caption{\label{signals-shotnoise}  Simulated sensor responses under an incident electromagnetic field with an amplitude $\Omega_{\rm{tg}} = 2\pi\times 1$ kHz (a-c) during the time interval $[0.5t_0, t_0]$ (a), $[2.5 t_0, 3t_0]$ (b) and $[9.5 t_0, 10t_0]$ (c), and with $\Omega_{\rm{tg}} = 2\pi\times 8.2$ kHz (d-e) during $[1.34, 1.41]$ ms,  labelled with   $'\ast'$ for $\xi = 2\pi\times0.12$ kHz (red), $'\circ'$ for $\xi = 2\pi\times0.06$ kHz (blue) and $'\times'$ for $\xi = 2\pi\times0.072$ kHz (black). Recall that $t_0=1.41$ ms. The solid lines represent the ideal $P_i$ (no shot noise) derived from solving the Schr\"{o}dinger equation and correspond to different $\xi$ in the respective color as mentioned above.	 
		(f) Simulated sensor responses for $\xi = 2\pi\times 0.06$ kHz  with $\Omega_{\rm{tg}} = 2\pi\times 1$ kHz ($'\times'$, black), $\Omega_{\rm{tg}} = 2\pi\times 1.1$ kHz ($'\circ'$, blue) and $\Omega_{\rm{tg}} = 2\pi\times 1.2$ kHz ($'\ast'$, red),  while the solid lines represent the sensor response ideal $P_i$ (no shot noise). In all the plots, the number of shots for each simulated measured sensor response is $N_m=100$ at each $t_i$.}
	\end{center}
\end{figure*}

\section{Impact of shot noise and evolution time for parameter detection}\label{app:b}
In order to illustrate the influence of shot noise on the measured sensor response, we show in Fig. \ref{signals-shotnoise} the ideal response (solid lines) together with a simulated experimental acquisition (labels) with $N_m=100$ measurements per time instance.

As an example, Fig.~\ref{signals-shotnoise} (a-c) indicate that, when $\Omega_{\rm{tg}}=2\pi \times 1$ kHz, the measured sensor response with $\xi=2\pi\times 0.12$ kHz and $\xi=2\pi\times 0.06$ kHz cannot be distinguished during the interval $[0.5t_0, t_0]$. Upon a longer evolution time, they become distinguishable, e.g. for $[2.5 t_0, 3t_0]$. However, until $[9.5 t_0, 10t_0]$ the sensor response with $\xi = 2\pi\times 0.06$ kHz and $\xi = 2\pi\times 0.072$ kHz are separated. 
In contrast, when $\Omega_{\rm{tg}}$ gets larger, the sensor response obtained from detunings with small difference are easier to be recognized at shorter times. For instance, 
as shown in Fig.~\ref{signals-shotnoise} (d-e), at $\Omega_{\rm{tg}} = 2\pi\times8.2$ kHz ($\Omega_{\rm{tg}}^r = 0.3$), during $[1.34, 1.41] $ ms, the data points measured from $\xi=2\pi\times0.06$ kHz almost overlap with the ones from $\xi = 2\pi\times 0.072$ kHz. Nevertheless, they can be differentiated from the ones with $\xi = 2\pi\times 0.12$ kHz. In Fig.~\ref{signals-shotnoise} (f), given by the same $\xi = 2\pi\times0.06$ kHz, a small change in the Rabi frequency leads to a big variation in the response during the time interval $[0.5t_0, t_0]$, even in the regime $\Omega_{\rm{tg}} \approx 2\pi\times1$ kHz.

\end{document}